\documentclass[journal=jacsat,manuscript=article]{achemso}

\usepackage[version=3]{mhchem} 

\usepackage{color}
\usepackage{gensymb}

\author{Oleh Y. Yermakov}
\affiliation{Department of Nanophotonics and Metamaterials, ITMO University, St. Petersburg 197101, Russia}
\email{oe.yermakov@gmail.com}
\author{Dmitry V. Permyakov}
\affiliation{Department of Nanophotonics and Metamaterials, ITMO University, St. Petersburg 197101, Russia}
\author{Filipp V. Porubaev}
\affiliation{Department of Biblical Studies, St. Petersburg Theological Academy, St. Petersburg 191167, Russia}
\author{Pavel~A.~Dmitriev}
\affiliation{Department of Nanophotonics and Metamaterials, ITMO University, St. Petersburg 197101, Russia}
\author{Dmitry A. Baranov}
\affiliation{Department of Nanophotonics and Metamaterials, ITMO University, St. Petersburg 197101, Russia}
\author{Anton~K. Samusev}
\affiliation{Department of Nanophotonics and Metamaterials, ITMO University, St. Petersburg 197101, Russia}
\author{Ivan V. Iorsh}
\affiliation{Department of Nanophotonics and Metamaterials, ITMO University, St. Petersburg 197101, Russia}
\author{Radu Malureanu}
\affiliation{DTU Fotonik, Technical University of Denmark, Oersteds pl. 343, DK-2800 Kongens Lyngby, Denmark}
\author{Andrey A. Bogdanov}
\email{bogdan.taurus@gmail.com}
\affiliation{Department of Nanophotonics and Metamaterials, ITMO University, St. Petersburg 197101, Russia}
\author{Andrei V. Lavrinenko}
\affiliation{Department of Nanophotonics and Metamaterials, ITMO University, St. Petersburg 197101, Russia}
\alsoaffiliation{DTU Fotonik, Technical University of Denmark, Oersteds pl. 343, DK-2800 Kongens Lyngby, Denmark}

\title[ACS_paper]
  {Effective surface conductivity of plasmonic metasurfaces: from far-field characterization to surface wave analysis}

\begin{document}


\begin{abstract}
Metasurfaces offer great potential to control near- and far-fields through engineering of optical properties of elementary cells or meta-atoms. Such perspective opens a route to efficient manipulation of the optical signals both at nanoscale and in photonics applications. In this paper we show that by using an effective surface conductivity tensor it is possible to unambigiously describe optical properties of an  anisotropic metasurface in the far- and near-field regimes. We begin with retrieving the effective surface conductivity tensor from the comparative analysis of experimental and numerical reflectance spectra of a metasurface composed of elliptical gold nanoparticles. Afterwards restored conductivities are validated in the crosscheck versus semianalytic parameters obtained with the discrete dipole model with and without dipoles interaction contribution. The obtained effective parameters are further used for the dispersion analysis of surface plasmons localized at the metasurface. The effective medium model predicts existence of both TE- and TM-polarized plasmons in a wide range of optical frequencies and describes peculiarities of their dispersion, in particularly, topological transition from the elliptical to hyperbolic regime with eligible accuracy. The analysis in question offers a simple practical way to describe properties of metasurfaces including ones in the near-field zone by extracting effective parameters from the convenient far-field characterisation. 

    
\end{abstract}

\section{Introduction}

Miniaturization of integrated optical circuits requires an effective control of light on the subwavelength scale. Significant advances in this field have been achieved with the help of metamaterials~\cite{smith2004metamaterials,engheta2006metamaterials, shalaev2007optical} -- artificially created media, whose electromagnetic properties can drastically differ from the properties of the natural materials. However, a three-dimensional structure of metamaterials, related fabrication challengers and high costs, especially for optical applications, form significant obstacles in their implementation in integrated optical circuits.

An alternative way is to use \textit{metasurfaces} -- two-dimensional analogues of metamaterials. There are also natural two-dimensional anisotropic materials such as hexagonal boron nitride~\cite{dai2015subdiffractional,li2015hyperbolic}, transition metal dichalcogenides~\cite{hamm2013two, glazov2014exciton}, black phosphorus~\cite{correas2016black}. In the visible and the near-IR range, metasurfaces can be implemented using  subwavelength periodic arrays of plasmonic or high-index dielectric nanoparticles~\cite{yu2014flat, meinzer2014plasmonic, kuznetsov2016optically}. A nanostructured graphene could also be considered as a metasurface for THz frequencies~\cite{christensen2011graphene,trushkov2015two}. In the microwave range, metasurfaces can be implemented by using LC-circuits, split-ring resonators, arrays of capacitive and inductive elements (strips, grids, mushrooms), wire medium etc~\cite{holloway2012overview,glybovski2016}. Despite subwavelength or even monoatomic thicknesses, the metasurfaces offer unprecedented control over light propagation, reflection and refraction \cite{holloway2012overview,yu2015optical}.

Metasurfaces exhibit a lot of intriguing properties for a wide area of applications such as near-field microscopy, imaging, holography, biosensing, photovoltaics etc~\cite{holloway2012overview,yu2014flat,yu2015optical,glybovski2016}. For instance, it was shown that metasurfaces based on Si nanoparticles can exhibit nearly 100\% reflectance~\cite{moitra2014experimental} and transmittance~\cite{decker2015high} in a broadband frequency range. Moreover, metasurfaces can serve as light control elements: frequency selectors, antennas, lenses, perfect absorbers \cite{glybovski2016}. They offer an excellent functionality with polarization conversion, beam shaping and optical vortices generation~\cite{yang2014dielectric,desiatov2015polarization}.  Besides, metasurfaces provide an efficient control over dispersion and polarization of surface waves \cite{takayama2017photonic,kildishev2013planar,yu2014flat,yermakov2015hybrid,gomez2015hyperbolic,low2017polaritons}. Surface plasmon-polaritons propagating along a metasurface assist pushing, pulling and lateral optical forces in its vicinity~\cite{rodriguez2015lateral, petrov2016surface}. Metasurfaces are prospective tools for spin-controlled optical phenomena \cite{shitrit2013spin, aiello2015transverse,bliokh2015transverse,yermakov2016spin} and holographic applications~\cite{huang2013three,ni2013metasurface,zheng2015metasurface}. The main advantages of metasurfaces, such as relative manufacturing simplicity, rich functionality and planar geometry, fully compatible with modern fabrication technologies, create a promising platform for the \textit{photonic metadevices}. It has been recently  pointed out that all-dielectric metasurfaces and metamaterials can serve as a prospective low loss platform, which could replace plasmonic structures~\cite{jahani2016all}. However, one of the main advantages of plasmonic structures unachievable with dielectric ones is that the plasmonic structures can be resonant in the visible range keeping at the same time a deep subwavelength thickness and period. Thus, here we concentrate on  plasmonic metasurfaces allowing light manipulation with a deep subwavelength structure.

The common feature of bulk metamaterials and metasurfaces is that due to the subwavelength structure they can be considered as homogenized media described by effective material parameters. For bulk metamaterials, such effective parameters are permittivity~$\hat{\varepsilon}_{\text{eff}}$ and/or permeability~$\hat{\mu}_{\text{eff}}$. Retrieving effective parameters is one of the most important problems in the study of metamaterials. Generally, the effective parameters are tensorial functions of frequency~$\omega$, wavevector~$\mathbf{k}$, and intensity $I$. Homogenization of micro- and nanostructured metamaterials can become rather cumbersome, especially taking into account nonlocality \cite {simovski2010electromagnetic,chebykin2012nonlocal}, chirality \cite{andryieuski2010homogenization}, bi-anisotropy \cite{ouchetto2006homogenization, alu2011first} and nonlinearity \cite{mackay2005linear,Larouche2010}.



Analogous homogenization procedures are relevant for metasurfaces. Apparently, homogenization procedures for 2D structures were firstly developed in radiophysics and microwaves (equivalent surface impedance) in applications to thin films, high-impedance surfaces and wire grids etc~\cite{macfarlane1946surface,klein1990effective,tretyakov2003thin}. It has  been recently pointed out that two-dimensional structures, like graphene, silicene and metasurfaces, can be described within an effective conductivity approach \cite{andryieuski2013graphene,tabert2013magneto,yermakov2015hybrid,danaeifar2015equivalent,gomez2015hyperbolic,nemilentsau2016anisotropic}. In virtue of a subwavelength thickness, a metasurface could be considered as a two-dimensional equvalent current and, therefore, characterized by effective electric $\hat\sigma_e(\omega, \mathbf{k_{\tau}})$ and magnetic $\hat\sigma_m(\omega, \mathbf{k_{\tau}})$ surface conductivity tensors, where $\mathbf{k_{\tau}}$ is the component of the wavevector in the plane of the metasurface \cite{holloway2012overview,glybovski2016}. Importantly, such effective surface conductivity describes the properties of the metasurface both in the far-field when $|\mathbf{k_{\tau}}| < \omega/c$ (reflection, absorption, refraction, polarization transformation etc.) and in the near-field (surface waves, Purcell effect, optical forces), when $|\mathbf{k_{\tau}}| > \omega/c$.



In this paper, we focus our study on a resonant plasmonic anisotropic metasurface represented by a two-dimensional periodic array of gold nanodisks with the elliptical base. We derive and analyze the electric surface conductivity tensor of the anisotropic metasurface in three ways: (i) theoretically by using the  discrete dipole model; (ii) experimentally by characterization of the metasurface reflection spectra and (iii) numerically by combining the optical measurements of the fabricated metasurface, simulations of the experiment and analytical approach (zero-thickness approximation). We reveal that the effective surface conductivity tensor extracted from the far-field measurements well describes near-field properties of metasurface such as the spectrum of surface waves and their behaviour in all possible regimes - capacitive, inductive, and hyperbolic. By using the discrete dipole model we study the effects of spatial dispersion on the eigenmodes spectrum and define the limitations of the effective model applicability.

\section{Sample Design and Fabrication}

\begin{figure}[t!]\centering
  \includegraphics[width=.78\textwidth]{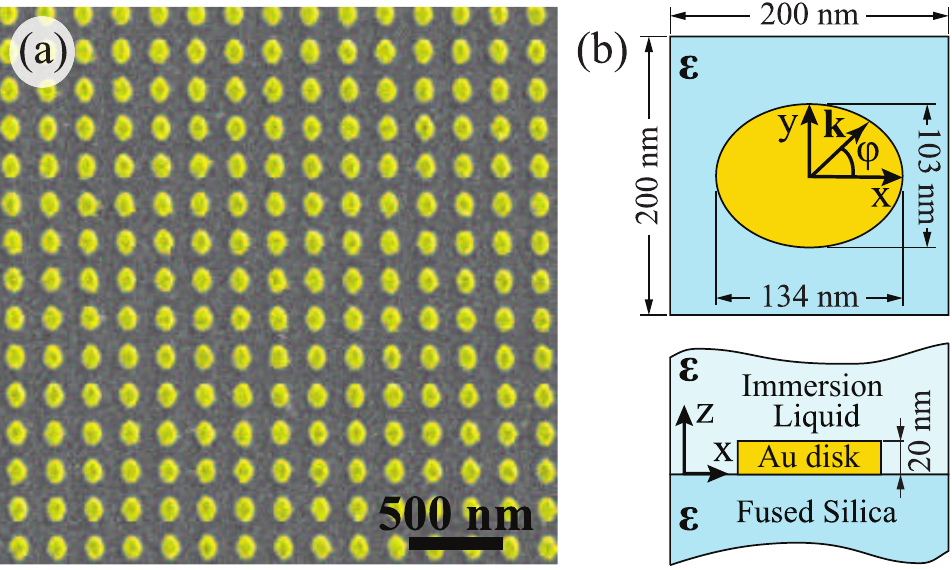}
  \caption{(a) False color SEM image of the fabricated anisotropic metasurface. b) The structure consists of 20-nm-thick gold nanodisks arranged in a square lattice (period 200~nm). The base of the disks has an elliptical shape, with the long and short axes equal to 134 and 103~nm, respectively. We assume the environment is uniform and isotropic with $\varepsilon = 2.1$. 
  }
 \label{fig:Design}
\end{figure}

We consider a metasurface composed of gold anisotropic nanoparticles placed on a fused silica substrate. The design of the sample is shown in Fig.~\ref{fig:Design}. The target structure consists of 20~nm thick gold nanodisks with the  elliptical base packed in the square lattice with a period of 200~nm. The average long and short axes of the disks are $a_x = 134$~nm and $a_y = 103$~nm, respectively. The distribution of the nanodisks sizes is provided in Fig.~6 (See Supporting Information). 

The sample was fabricated via electron beam lithography on a fused silica substrate. Before the electron beam exposure process, the resist layer (PMMA) was covered with a thin gold layer to prevent local charge accumulation. After the exposure, a 20~nm thick gold layer was sputtered via thermal evaporation. During the last step of the fabrication process, the remains of the resist were removed via the lift-off procedure. Finally, the sample was immersed in a liquid with a refractive index nearly matching the glass substrate. Thus, we obtained the metasurface with a homogeneous ambient medium with permittivity $\varepsilon = 2.1$. The SEM image of the fabricated sample is shown in Fig.~\ref{fig:Design}a.

\section{Effective Conductivity Tensor}
The plasmonic resonant metasurface shown in Fig. 1 is anisotropic and non-chiral. Asymmetry of each particle splits its in-plane dipole plasmonic resonance with frequency $\Omega$ into two resonances with frequencies $\Omega_\bot$  and $\Omega_\|$~\cite{yermakov2015hybrid, samusev2017two}. Consideration of metasurfaces as an absolutely flat object might be restricted due to the emergence of the out-of-plane polarizability caused by the finite thickness of the plasmonic particles. In our case, the out-of-plane polarizability $\alpha_{z}$ is neglected due to a small thickness of the particles as it is shown in Fig.~7 (See Supporting Information). Therefore, this metasurface can be described by a two-dimensional effective surface conductivity tensor diagonal in the principal axes (when the axes of coordinates systems are parallel to the axes of the elliptical base of the nanodisks).

\subsection{Zero-thickness Approximation}

To extract the effective surface conductivity of the fabricated sample, we apply a procedure based on the combination of the optical experiments, numerical simulations and theoretical calculations. 

\begin{figure}[t!]
\centering
  \includegraphics[width=.92\textwidth]{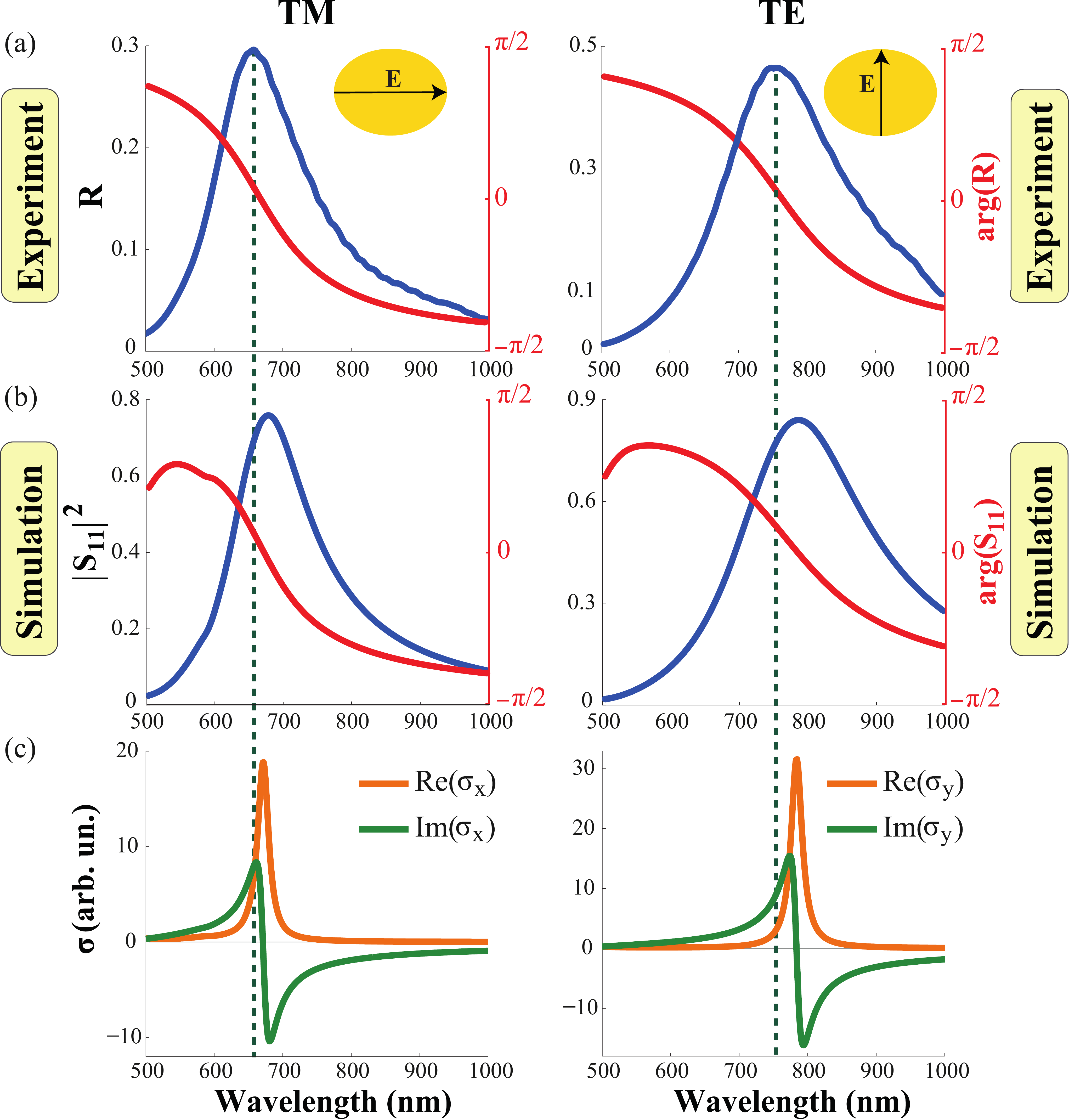}
  \caption{Reflectance spectra of a metasurface for polarization along (left panel) and across (right panel) the long axis of the disk. (a) Intensity (blue lines) and phase (red lines)  of the reflectance spectra obtained from the experimental measurements. (b) Squared moduli (blue lines) and phases (red lines) of the metasurface reflection coefficient $S_{11}$ calculated in CST Microwave Studio. (c) Real (orange lines) and imaginary (green lines) parts of the TM- and TE-polarized components of the effective surface conductivity tensor extracted from $S_{11}$ data via ZTA.}\label{fig:MetasurfaceData}
\end{figure}

First, we measure the intensity of the reflectance for the light polarized along and across the principle axes of the metasurface under normal incidence (Fig.~\ref{fig:MetasurfaceData}a). Both spectra demonstrate single peaks corresponding to the individual localized plasmon resonances of the nanodisks. The phase retrieved by the fitting of the experimental reflectance with the intensity calculated by the use of the Drude formula (See Supporting Information) is shown in Fig.~\ref{fig:MetasurfaceData}a by the red lines. 

Then, we model the experiment with CST Microwave Studio~(Fig.~\ref{fig:MetasurfaceData}b). The difference in the intensity of the peaks in Figs.~\ref{fig:MetasurfaceData}a and \ref{fig:MetasurfaceData}b can be attributed to roughness and inhomogeneity of the sample. To equalize the measured and simulated peak values We increase the imaginary part of the gold permittivity in the simulations (See Supporting Information). Retrieving the complex conductivity tensor is done by applying the intensity and phase of the modeled reflectance, wherein we obtain a good matching between the simulated and experimental shapes of the reflectance spectra~(Figs.~\ref{fig:MetasurfaceData}a and ~\ref{fig:MetasurfaceData}b).

Basing on the calculated complex reflection coefficient we find an effective surface conductivity using the zero-thickness approximation (ZTA). Within this approximation we replace the real structure of finite thickness $H$ by the effective two-dimensional plane disposed at distance $H/2$ from the substrate. This technique can be applied only for deeply subwavelength structures. The limitation can be formulated as $H/\lambda \ll 1$ according to the Nicholson-Ross-Weir method~\cite{baker1990improved,luukkonen2011stepwise}.

Considering a two-dimensional layer with effective conductivity $\sigma$ sandwiched between two media with refractive indices $n_1$ (superstrate) and $n_2$ (substrate) one can find Fresnel's coefficients \cite{andryieuski2013graphene,merano2016fresnel} and express the effective surface conductivity as follows
\begin{equation}
\sigma_{x,y} = \frac{n_1 - n_2 - S_{11}^{x,y}(n_1+n_2)}{1+S_{11}^{x,y}},
\label{sigma}
\end{equation}
where $S_{11}^{x,y}$ is the component of the $S$-matrix. Indices $x,y$ correspond to different orientations of the electric field of the incident wave. Hereinafter we use the Gauss system of units and express surface conductivity in the dimensionless units $\sigma = 4 \pi \widetilde{\sigma}/c$.

In order to obtain the proper conductivity of a metasurface one should retain only the phase of the reflection coefficient related to the metasurface properties. In the simulation, the total phase of the $S$-parameters has two contributions $\text{arg}(S_{11}^{x,y})=\Delta\varphi_1+\Delta\varphi_2$. The first one arises directly when the wave reflects from the metasurface.  The second phase arises because of the wave propagation from the port to the metasurface and back $\Delta\varphi_2=2k_0L$\footnote{The time dependence is defined through the factor $e^{i \omega t}$.}. Here $k_0 = n_1 \omega/c$, $L$ is the distance between the excitation port and the metasurface. The problem is how to correctly determine distance $L$ if the metasurface has a finite thickness? We found that the correct results not breaking the energy conservation law (see Supporting Information) are obtained only if $L$ is defined as the distance to the middle of the metasurface. Thus, the effective two-dimensional layer has to be disposed exactly at  distance $H/2$ from the substrate. The obvious analogue of ZTA is the transfer matrix method (TMM), which originates from Fresnel's reflection and transmission coefficients. For the metasurface under consideration ZTA and TMM give the results with the average relative error of 1\%. However, the advantage of ZTA over TMM is that it is necessary to know only one either reflection or transmission coefficient to extract the effective parameters. 


Extracted conductivities for both polarizations are presented in Fig.~\ref{fig:MetasurfaceData}c. For the light wave polarized along the long axis (TM-polarization) the plasmon resonance is observed at 670 nm, while for light polarized along the short axis (TE-polarization) the resonance corresponds to 780~nm.

\subsection{Discrete Dipole Model}

In order to derive surface conductivity of a metasurface analytically we apply the \textit{discrete dipole model} (DDM)\footnote{In many works it is also called the \textit{point-dipole model}.}. This technique has been implemented for 1D, 2D and 3D structures~\cite{moroz2001exponentially, lunnemann2014dispersion, belov2005homogenization, poddubny2012microscopic,chebykin2015spatial}. Within this approach we consider a 2D periodic array of the identical scatterers as an array of point dipoles. 


In the framework of the DDM it is more convenient to operate with an effective polarizability of the structure, which is straightforwardly connected to the effective conductivity tensor as follows:
\begin{equation}
\hat{\sigma}_\text{eff} = -\frac{i 4 \pi \varepsilon \omega}{c a^2} \hat{\alpha}_\text{eff}.
\end{equation}
In the case under consideration, the thickness of the scatterers is deeply subwavelength and, therefore, we can neglect the polarizability of the particles in the direction perpendicular to the plane of the metasurface. Thus, we can describe the metasurface by either two-dimensional polarizability tensor $\hat{\alpha}_{\text{eff}}$ or conductivity tensor $\hat{\sigma}_{\text{eff}}$  with zero off-diagonal components (in the basis of the principal axes). Rigorous derivation of the effective polarizability of a two-dimensional lattice of resonant scatterers is performed in  Refs.~\citenum{tretyakov2003impedance, alu2011first,belov2005homogenization}. The effective polarizability of the metasurface can be written as
\begin{align}
\hat{\alpha}_{\text{eff}}^{-1} (\omega, \mathbf{k_{\tau}})=\hat{\alpha}_0^{-1}(\omega) + \hat{C}(\omega, \mathbf{k_{\tau}}) \label{alphafull}.
\end{align}
Here, $\hat{\alpha}_0(\omega)$ is the polarizability of the individual resonant scatterer, and $\hat{C}(\omega, \mathbf{k_{\tau}})$ is the so-called dynamic interaction constant~\cite{belov2005homogenization}. The latter contains the \textit{lattice sum}, which takes into account interaction of each dipole with all others. We approximate the polarizability of the disk with the elliptical base $\hat{\alpha}_0$ by the polarizability of an ellipsoid with the same volume and aspect ratio (See Supporting Information). We calculate the interaction between the identical scatterers by using the Green's function formalism:
\begin{equation}
\hat{C}(\omega, \mathbf{k_{\tau}}) = \sum_{i,j} \hat{G}(\omega, {\bf r}_{ij}) e^{i {{\bf k_{\tau}} {\bf r}_{ij}}}.
\label{alpha_green}
\end{equation}
Here $\hat{G}(\omega,{\bf r}_{ij})$ is the dyadic Green's function and ${\bf r}_{ij}$ are the coordinates of the dipoles. This sum has slow convergence. So, we calculate the interaction term in Eq.~\eqref{alpha_green} within the Ewald summation technique~\cite{moroz2001exponentially,silveirinha2005new,capolino2007efficient,poddubny2012microscopic} applied for a two-dimensional periodic structure, which ensures fast convergence of the sum (See Supporting Information).

The discrete dipole model can be successfully applied for many types of metasurfaces. It is applicable for two-dimensional periodic structures under three main conditions:
\begin{itemize}
\item[1.] Quasistatic condition: $n a \ll \lambda$. Here $n$ is the refractive index of the environment, $a$ is the lattice constant, $\lambda$ is the incident wavelength.
\item[2.] Dipole approximation: $f \ll 1$ (or $d \ll a$, where $d$ is the characteristic size of a scaterrer). Here $f = A/a^2$ is the filling factor, $A$ is the area occupied by the scatterer in the unit cell (in our case, $A = \pi a_x a_y$), $a^2$ is the area of the square unit cell. When the scatterers are not sufficiently small one has to take into account higher order multipoles.
\item[3.] Quasi-two-dimensionality: $\alpha_{zz} \ll \text{min}\left( \alpha_{xx}, \alpha_{yy} \right)$ and $H \ll \lambda$. This condition is achieved, when thickness of a metasurface is less than both characteristic in-plane sizes of meta-atoms ($H < \text{min}\left\lbrace a_x, a_y \right\rbrace$) and skin depth $\delta$ ($H < \delta$). 
\end{itemize}
For the metasurface sample under consideration $H/a_y = 0.19$, and $f=0.28$. Although the applicability condition of the dipole approximation is poorly satisfied, the DDM gives eligible results. Parameters $n a/\lambda$ and $H/\lambda$ lie in the interval from 0.25 to 0.75 and from 0.02 to 0.05, respectively,  for wavelengths $\lambda=400-1200$~nm. Skin depth $\delta$ for gold is around 20-40~nm in the wavelength range under consideration~\cite{olmon2012optical}.

\begin{figure}[htbp]\centering
  \includegraphics[width=.86\textwidth]{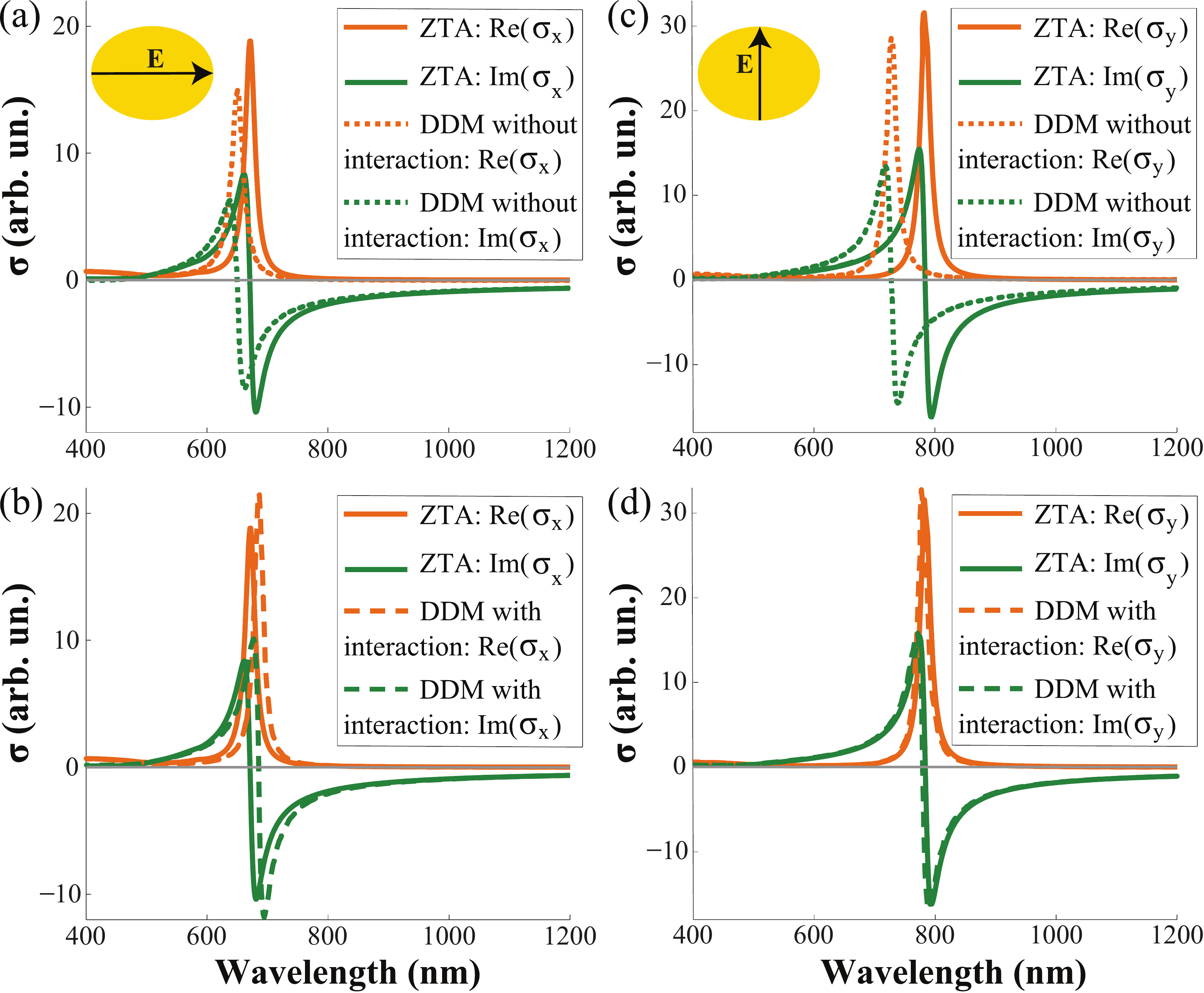}
  \caption{Real (orange lines) and imaginary (green lines) parts of the effective conductivity extracted via ZTA (solid lines), DDM without interaction (dotted lines) and DDM with interaction (dashed lines)  for TM (a,b) and TE (c,d) polarizations.}
\end{figure}

One can see in Figs.~3a and 3c that neglecting interaction term $\hat{C}(\omega,\mathbf{k_{\tau}})$ in Eq.~\eqref{alphafull} results in a blue shift of the conductivity spectra  by several tens of nanometers for both polarizations. Accounting these interactions brings the DDM into almost perfect agreement with the ZTA (Figs.~3b and 3d). However,  matching for $\sigma_y$ is better than for $\sigma_x$. It could be explained by the fact that polarizability of an ellipsoid approximates polarizability of the elliptical disk in the $y$ direction better that in the $x$ direction.


\subsection{Analysis}

The spectral dependences of the extracted surface conductivities along the principal axes are shown in Fig.~\ref{fig:MetasurfaceData}c and Fig.~3. They clearly show that the fabricated metasurface is characterized by a highly anisotropic resonant conductivity tensor:
\begin{equation}
\hat{\sigma}_{\text{eff}}= \left(
\begin{matrix}
\sigma_{x} & 0  \\
0 & \sigma_{y}
\end{matrix}
\right).
\label{sigma_tensor}
\end{equation}
One can see from Fig.~\ref{fig:MetasurfaceData}c that the metasurface supports three different regimes depending on  wavelength $\lambda$ of the incident light. These regimes can be classified by the signs of (i) $\text{det}[\text{Im}(\hat\sigma)]$ and (ii) $\text{tr}[\text{Im}(\hat\sigma)]$. Specifically, when  $\text{det}[\text{Im}(\hat\sigma)] > 0$ and $\text{tr}[\text{Im}(\hat\sigma)] > 0$ (for $\lambda <$ 670 nm) the inductive regime of the metasurface is observed. In this case, the metasurface corresponds to the conventional metal sheet and only a TM-polarized surface wave can propagate. For $\text{det}[\text{Im}(\hat\sigma)] > 0$ and $\text{tr}[\text{Im}(\hat\sigma)] < 0$ (for $\lambda >$ 780 nm) the capacitive regime of the metasurface is met, so only a TE-polarized surface wave can propagate. When $\text{det}[\text{Im}(\hat\sigma)] < 0$ (between the resonances, i.e. for wavelengths from 670 to 780 nm), a metasurface supports the so-called \textit{hyperbolic} regime, in which simultaneous propagation of both TE- and TM-modes is possible~\cite{yermakov2015hybrid}. 

\section{Surface Waves}

In this Section, we analyze the spectrum of the surface waves supported by the metasurface using the extracted effective conductivity tensor and compare the results with full-wave numerical simulations. 

The dispersion equation of the surface waves supported by an anisotropic metasurface, described by the effective conductivity tensor \eqref{sigma_tensor}, can be straightforwardly derived from Maxwell's equations and boundary conditions at the metasurface~\cite{yermakov2015hybrid}:
\begin{equation}
\begin{split}
\label{disp_eq}
\left( \frac{\kappa_1}{ \mu_1 k_0} + \frac{\kappa_2}{\mu_2 k_0} - i\sigma_{xx}\right) \left( \frac{\varepsilon_1 k_0}{\kappa_1} + \frac{\varepsilon_2 k_0}{\kappa_2} + i\sigma_{yy} \right) = \sigma_{xy}\sigma_{yx}.
\end{split}
\end{equation}
Here, $\sigma_{ij}$ are the tensor components in the coordinate system rotated by angle $\varphi$ (see Fig.~1b), $\varepsilon_1,\mu_1,\kappa_{1}$ and $\varepsilon_2,\mu_2,\kappa_{2}$ are the permittivity, permeability and inverse penetration depths of the wave in the superstrate and substrate, respectively. The latter are defined as $\kappa_i = \sqrt{\mathbf{k_{\tau}}^2 - \varepsilon_i \mu_i \omega^2/c^2}$, where $\mathbf{k_{\tau}}$ is the wavevector in the plane of the metasurface. In our case Eq.~(\ref{disp_eq}) is simplified since we consider the metasurface in non-magnetic ($\mu_1 = \mu_2 = 1$) and homogeneous environment with the permittivity corresponding to fused silica $\varepsilon = \varepsilon_1 = \varepsilon_2 = 2.1$.

The first and the second factors in the left side of Eq.~(\ref{disp_eq}) correspond to the dispersion of purely TE-polarized and TM-polarized surface waves, respectively. The right side of Eq.~(\ref{disp_eq}) is the coupling factor responsible for the mixing of TE and TM modes. If an electromagnetic wave propagates along a principal axis the coupling factor is zero, so either a conventional TM-plasmon or TE-plasmon exists. However, due to anisotropy ($\varphi \neq 0^{\circ}$) the coupling factor can become non-zero giving rise to hybrid surface waves of mixed TE-TM polarizations. Despite the hybridization, only one type of polarization is predominant for each mode. Therefore, it is logical to refer to such modes as \textit{quasi-TM} and \textit{quasi-TE} surface plasmons.


It is important to note that for a number of practical problems it is necessary to take into account nonlocal effects caused by spatial dispersion. Unfortunately, it can not be accounted for in the framework of the effective surface conductivity extracted from the normal incidence measurements. However, it can be calculated by using lattice sums. In this case, the dispersion equation for the eigenmodes has the following form: 
\begin{equation}
\text{det}|\hat{\alpha}_{\text{eff}}^{-1}(\omega, {\bf k_{\tau}})| = \text{det}|\hat{\alpha}_0^{-1}(\omega) +  \hat{C}(\omega, \mathbf{k_{\tau}})| = 0.
\label{disp_eq_lat_sum}
\end{equation}
Equation~\eqref{disp_eq_lat_sum} can be transformed into Eq.~\eqref{disp_eq} under the assumption that $d \ll a \ll \lambda$.

\begin{figure}[t!]\centering
  \includegraphics[width=.88\textwidth]{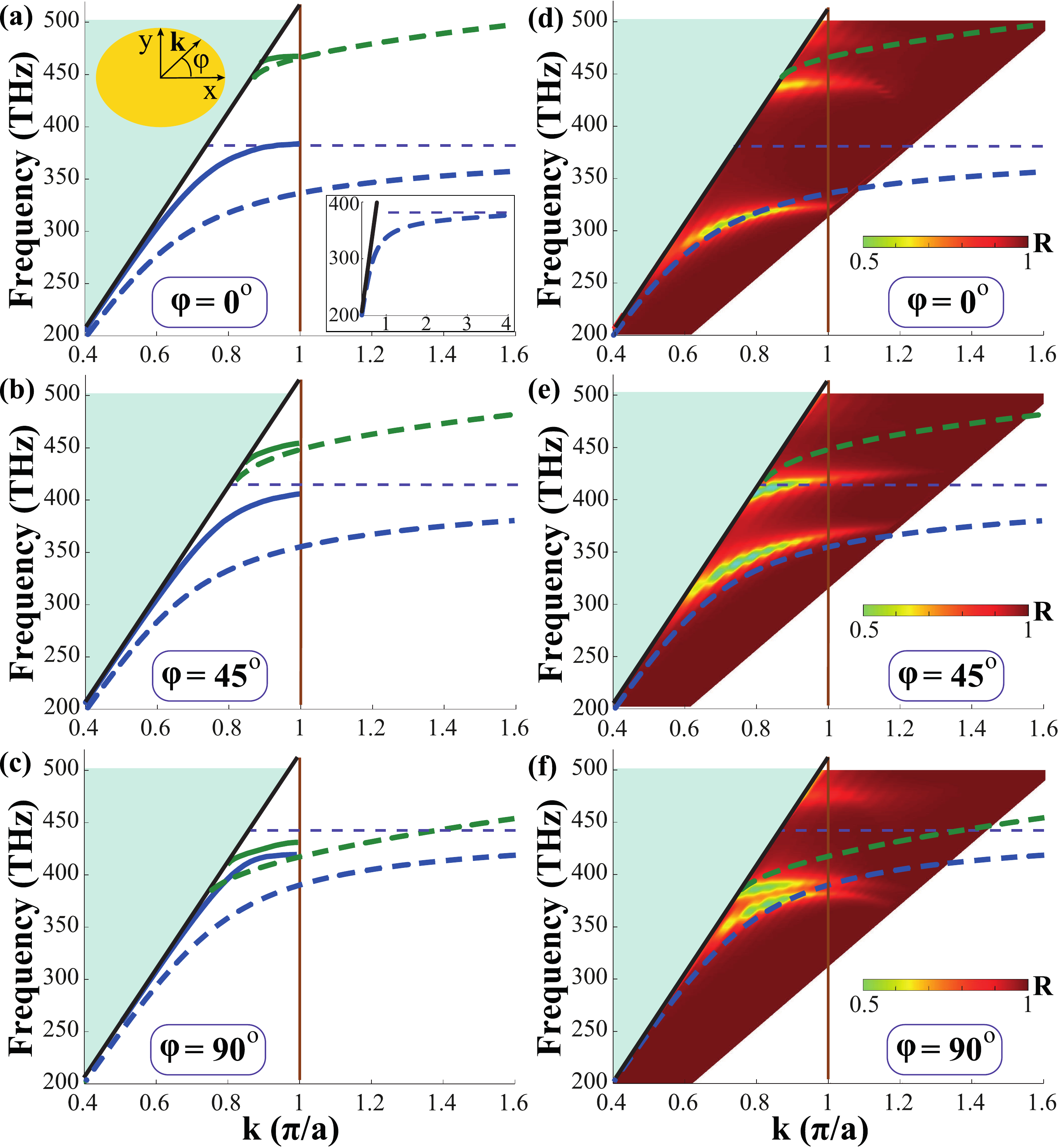}
  \caption{Dispersion of the quasi-TE (blue lines) and the quasi-TM (green lines) surface plasmons for different propagation angles (a,d) $\varphi = 0\degree$, (b,e) $\varphi = 45^{\circ}$, (c,f) $\varphi = 90^{\circ}$. Dashed lines correspond to the dispersion calculated within the effective conductivity approximation. Solid lines correspond to the calculations taking into account nonlocal interactions via the lattice sums (a-c). Color map shows the reflectance spectra of the metasurface coupled to a high-index prism with the mixed TE-TM polarization  (d-f). The light line in the medium with permittivity $\varepsilon = 2.1$ is shown as the black line. The boundary of the first Brillouin zone is shown as the vertical solid line. The horizontal purple dashed line corresponds to the plasmon resonance according to the effective model as the insert in sub-picture (a) shows.}
\end{figure}

Figure 4 shows the dispersion of the surface waves localized at the studied metasurface sample for different propagation angles $\varphi = 0, 45, 90^{\circ}$. In Figs.~4a-4c we compare the effective model and the discrete dipole model taking into account spatial dispersion ($\bf{k_{\tau}} \neq 0$). One can see that the difference in the dispersions obtained within the local and nonlocal models is significant. It can be explained by quite a large filling factor $f$, which sharply limits the accounting for nonlocal effects in the framework of the discrete dipole model. Nevertheless, both models are qualitatively similar. For instance, the resonant frequencies are close in both models for all propagation angles. Both models predict the frequency gap between TM- and TE-plasmons for $\varphi=0^{\circ}$ which shrinks with increasing of $\varphi$. At $\varphi=90^{\circ}$, the gap disappears and both surface modes can propagate at the same frequency, that is in accordance with the results of full-wave numerical simulations (see Fig.~4f). Better matching between the results of DDM and full-wave simulations could be obtained if we account for anisotropy of the dynamic interaction constant, but this theoretical extension is the subject of our further research.

To check the applicability of the effective conductivities extracted from the far-field measurements in characterization of the near-field phenomena, we compare dispersion of the surface waves from Figs.~4a-4c with the results from full-wave numerical simulations carried out in COMSOL Multiphysics (Figs.~4d-4e). One can see good correspondence of bands at low frequencies (for the quasi-TE mode). At high frequencies, i.e. small wavelengths, the effective model works worse but it is still eligible for qualitative results. 

\begin{figure}[htbp]\centering
  \includegraphics[width=.98\linewidth]{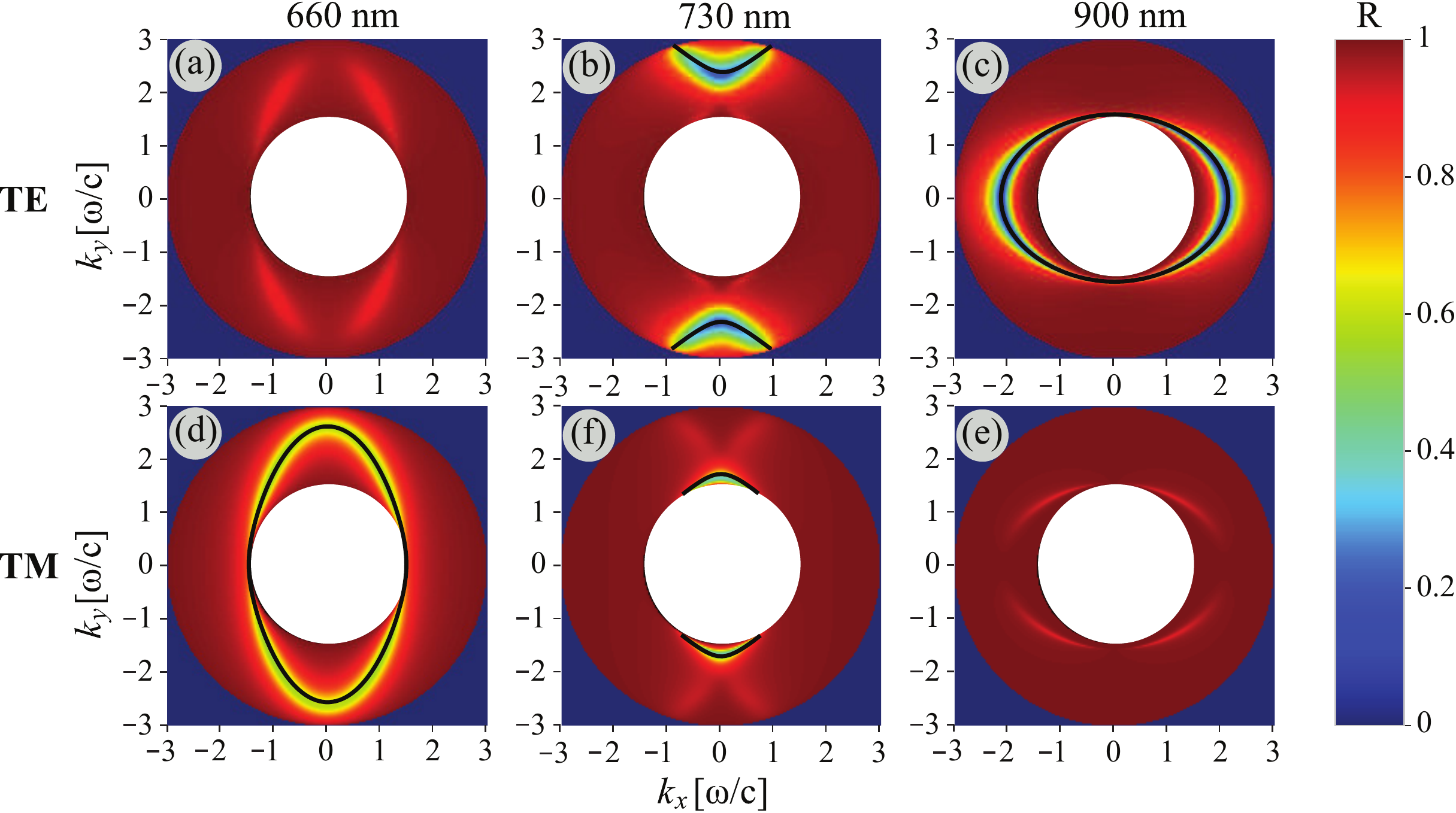}
  \caption{Simulation of the reflectance spectra from a metasurface shown in Fig.~\ref{fig:Design} for incident TE (a-c) and TM (d-e) polarizations. Panels (a) and (d), (b) and (f), (c) and (e) correspond to wavelengths $\lambda= 660, 730, 900$~nm, respectively. Black lines correspond to the equal frequency contours calculated from Eq.~\eqref{disp_eq} straightforwardly.}
 \label{fig:isofreq}
\end{figure}

It is convenient to present dispersion of surface waves in terms of equal frequency contours, which can be visualized in reflection experiments with a high index prism (Otto geometry). We calculate reflection of a light wave in such configuration by the transfer matrix method~\cite{pavel_dmitriev_2017_1041040}. When $\text{det}[\text{Im}(\hat\sigma)] > 0$, the equal frequency contours have an elliptic shape (Figs.~\ref{fig:isofreq}a,~\ref{fig:isofreq}c,~\ref{fig:isofreq}d,~\ref{fig:isofreq}e). For a hyperbolic regime, when $\text{det}[\text{Im}(\hat\sigma)] < 0$ ($\lambda = 730$ nm), the equal frequency contours represent a set of hyperbolas for the quasi-TE mode (Fig.~\ref{fig:isofreq}b) and arcs for the quasi-TM mode (Fig.~\ref{fig:isofreq}f). This drastic change of the shape is often called {\it topological transition}. One can see that in the hyperbolic regime both quasi-TE and quasi-TM modes are present, i.e. simultaneous propagation of two types of surface plasmons is observed (Figs.~\ref{fig:isofreq}b and ~\ref{fig:isofreq}f), which is consistent with bands dispersion in Fig.~4c and 4f. For the capacitive and inductive regimes only a single mode propagates. However, each mode has hybrid TE-TM polarization, so it is observed in both polarizations as shown in Fig.~5. Although polarization of the surface mode at 660 nm is predominantly similar to polarization of a conventional TM-plasmon (Fig.~5d), TE-polarization is also visible (Fig.~5a). The opposite situation takes place for a quasi-TE plasmon at $\lambda = 900$ nm (Figs.~5c and 5e). The exceptions are the principal axes directions where polarization of surface modes is strictly either purely TE or purely TM due to the lack of anisotropy.   

\section{Conclusions}

To conclude, we have suggested a practical concept to describe the full set of optical properties of a metasurface. Our approach is based on extraction of the effective surface conductivity. It allows to study various phenomena in the far-field as well as to calculate the spectrum of surface waves. We have developed two techniques to retrieve the effective conductivity and discussed their limitations. There are three different regimes of the local diagonal conductivity tensor of the anisotropic metasurface composed of elliptical gold nanodisks: inductive (metal-like), capacitive (dielectric-like) and hyperbolic (like in an indefinite medium). In contrast to an isotropic metasurface such anisotropic metasurface supports two modes of hybrid polarizations. We have shown the influence of non-locality on dispersion of the surface waves. Finally, we have demonstrated the topological transition of the equal frequency contours and the hybridization of two eigenmodes. We believe these results will be highly useful for a plethora of metasurfaces applications in nanophotonics, plasmonics, sensing and opto-electronics.

\begin{acknowledgement}

 This work was partially supported by the Villum Fonden, Denmark through the DarkSILD project (No. 11116), the the Ministry of Education and Science of the Russian Federation (3.1668.2017/4.6), RFBR (17-02-01234, 16-37-60064, 16-32-60123) and the Grant of the President of the Russian Federation (MK-403.2018.2). O.Y. acknowledges the support of the Foundation for the Advancement of Theoretical Physics and Mathematics “BASIS” (N\textsuperscript{\underline{o}} 17-15-604-1).

\end{acknowledgement}

\begin{suppinfo}

\section{Distribution of nanodisks sizes}

The fabrication of a metasurface is still challenged and complicated technological process. Obviously, not all particles have identical parameters, i.e. there is a distribution of particles position and sizes. According to such a distribution shown in Fig.~6 we define the average values of the long and short axes of the elliptical nanodisks bases as $a_x = 134.06 \pm 10.22$ and $a_y = 103.05 \pm 4.54$ nm, respectively. 

\begin{figure}[htbp]\centering
  \includegraphics[width=.75\linewidth]{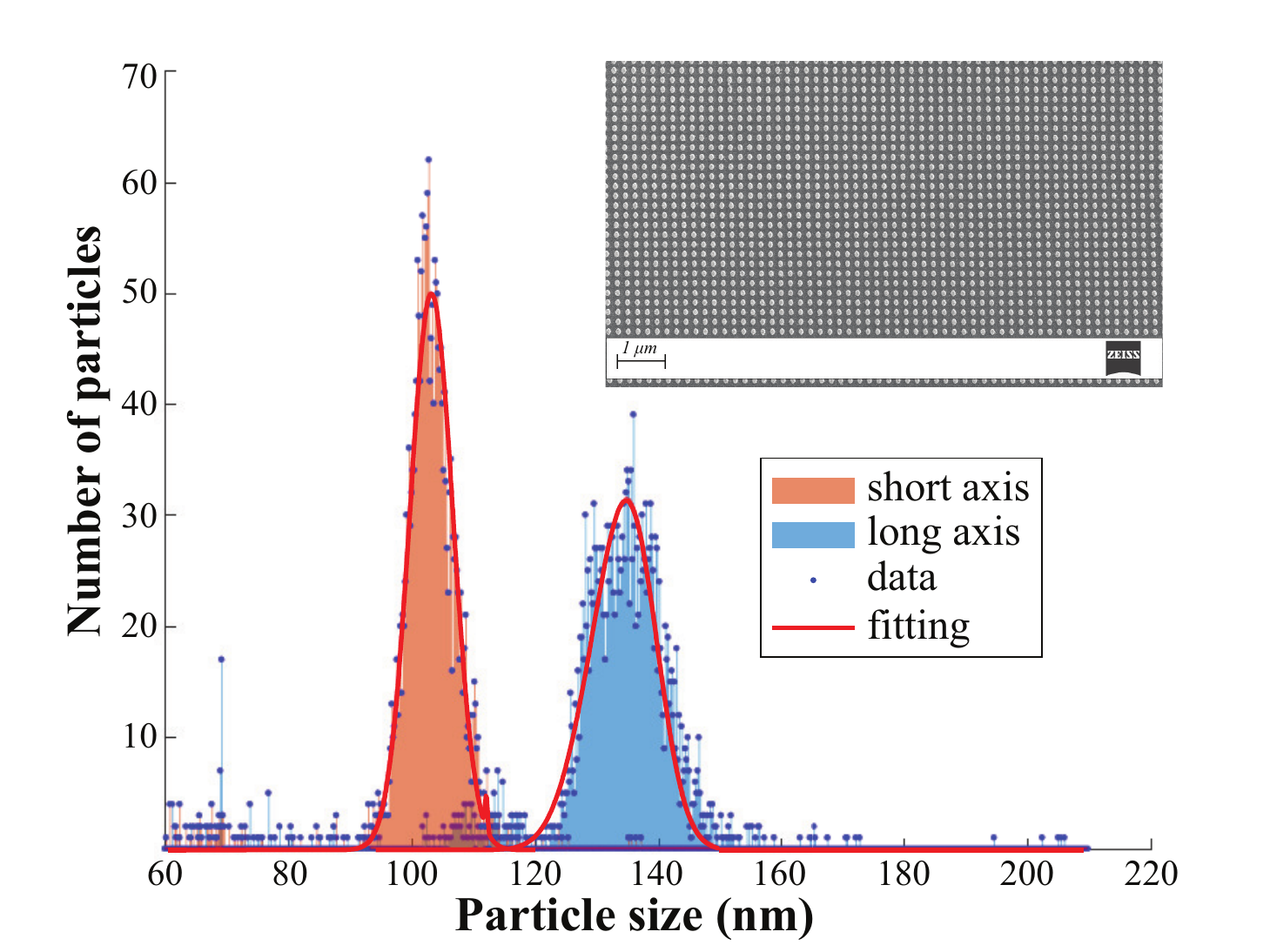}
  \caption{The distribution of the long and short axes of the elliptical nanodisks bases. The insert shows the SEM image of the fabricated metasurface sample.}
\end{figure}

\section{Polarizability of thin nanodisk with elliptical base}

We define the polarizability of a nanodisk with elliptical base $\hat{\alpha}_0 = \text{diag}(\alpha_x, \alpha_y, \alpha_z)$  through the polarizability of an ellipsoid. First, we consider the case of an ellipsoid with semiaxes $b_x$,~$b_y$,~$b_z$. For an anisotropic particle the depolarization factor $N_i$ should be introduced~\cite{landau2013electrodynamics}:
\begin{equation}
N_i = \frac{b_x b_y b_z}{2} \int_0^{\infty} \frac{ds}{\left( s + b_i^2 \right) \sqrt{(s + b_x^2)(s + b_y^2)(s + b_z^2)}}, \; i = x,y,z.
\tag{s1}
\end{equation}
Three depolarization factors for any ellipsoid satisfy the relation: $N_x + N_y + N_z = 1$. Finally, the polarizability of the ellipsoid is 
\begin{equation}
\alpha_{i} = \frac{b_x b_y b_z}{3} \frac{\varepsilon(\omega) - \varepsilon_m}{\varepsilon_m + N_i [\varepsilon(\omega) - \varepsilon_m]}, \; i = x,y,z.
\tag{s2}
\label{polarizability}
\end{equation}
Here $\varepsilon_m$ is the permittivity of the surrounding medium and $\varepsilon(\omega)$ is the permittivity of the scatterer material. In case of a sphere, when $b_x = b_y = b_z$, the depolarization factor is $N_i = 1/3$ and we get the polarizability of a sphere according to Clausius-Mossotti relation~\cite{jackson2007classical}. 

Then, we switch from the ellipsoid to the elliptical nanodisk with sizes $a_x, a_y, a_z = H/2$ and make the substitution for the semiaxes $b_i = (1.5)^{1/3} a_i$, that takes into account the difference between the volumes of an ellipsoid and an elliptical cylinder. After that we use Eq.~\eqref{polarizability} as the polarizability of the elliptical cylinder.

Thus, we obtain the polarizability of a thin nanodisk with elliptical base (Fig.~7). We conclude that resonance of the normal component of polarizability $\alpha_z$ is on very small wavelengths and for the studied range $\alpha_z$ is much smaller that in-plane components of polarizability. So, we can consider effective polarizability of thin nanodisk with elliptical base as a two-dimensional polarizability or conductivity tensor (5). 

\begin{figure}[htbp]\centering
  \includegraphics[width=.65\linewidth]{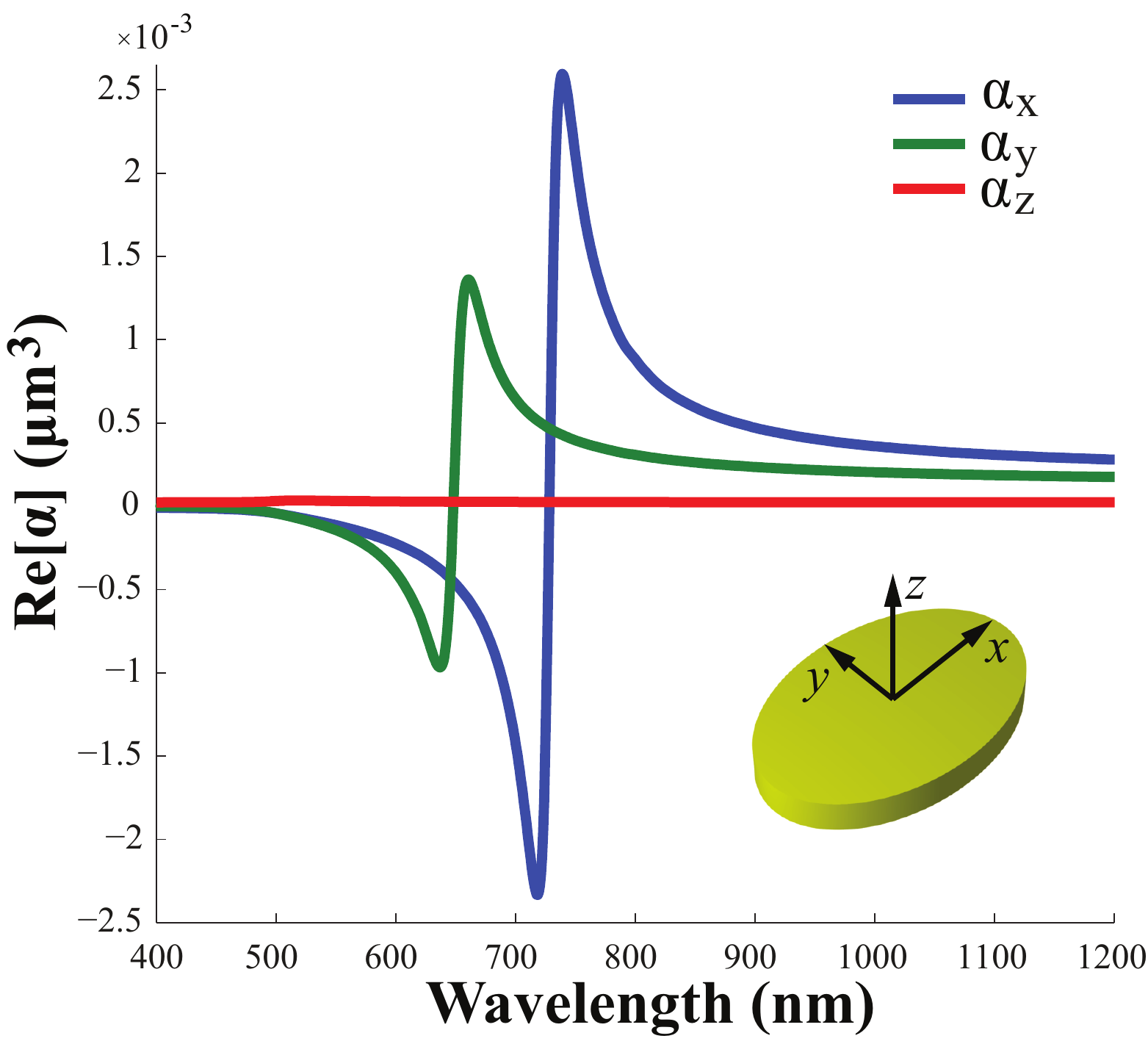}
  \caption{Real parts of the polarizability components for the thin nanodisk with elliptical base.}
\end{figure}

\section{Extraction of conductivity dispersion of two-dimensional layer by using fitting with Drude-Lorentz formula}

We introduce the Drude-Lorentz model with three undefined coefficients $\beta$, $\gamma$, $\sigma_0$:
\begin{equation}
\sigma = \frac{i \sigma_0}{\beta + i\gamma},
\label{sigma_drude}
\tag{s3}
\end{equation}
where $\sigma_0$ is the amplitude of conductivity dispersion, $\beta = \omega -\Omega^2/\omega$, $\omega$ is the operating frequency, $\Omega$ is the spectral position of resonance, $\gamma$ is the bandwidth of resonance.

Then we use Fresnel equations for a two-dimensional layer with an effective conductivity $\sigma$ located between two isotropic media $n_1$ and $n_2$ in order to express the reflection coefficient:
\begin{equation}
r = \frac{n_1 - n_2 - \sigma}{n_1 + n_2 + \sigma}.
\label{reflection}
\tag{s4}
\end{equation}
Here we use Gauss units and express surface conductivity in the normalized dimensionless units $\sigma = 4 \pi \widetilde{\sigma}/c$.

Substituting Eq.~\eqref{sigma_drude} into Eq.~\eqref{reflection} we obtain
\begin{equation}
r = \frac{\beta (n_1 - n_2) + i \left[ \gamma (n_1 - n_2) - \sigma_0  \right]}{\beta (n_1 + n_2) + i \left[ \gamma (n_1 + n_2) + \sigma_0  \right]},
\tag{s5}
\end{equation}
or expressing it through the reflectance
\begin{equation}
R = |r|^2 = \frac{\sigma_0^2 + (\beta^2 + \gamma^2)(n_1 - n_2)^2 - 2 \gamma \sigma_0 (n_1 - n_2) }{\sigma_0^2 + (\beta^2 + \gamma^2)(n_1 + n_2)^2 + 2 \gamma \sigma_0 (n_1 + n_2)}.
\label{fitting_formula}
\tag{s6}
\end{equation}

\begin{figure}[htbp]\centering
  \includegraphics[width=.88\linewidth]{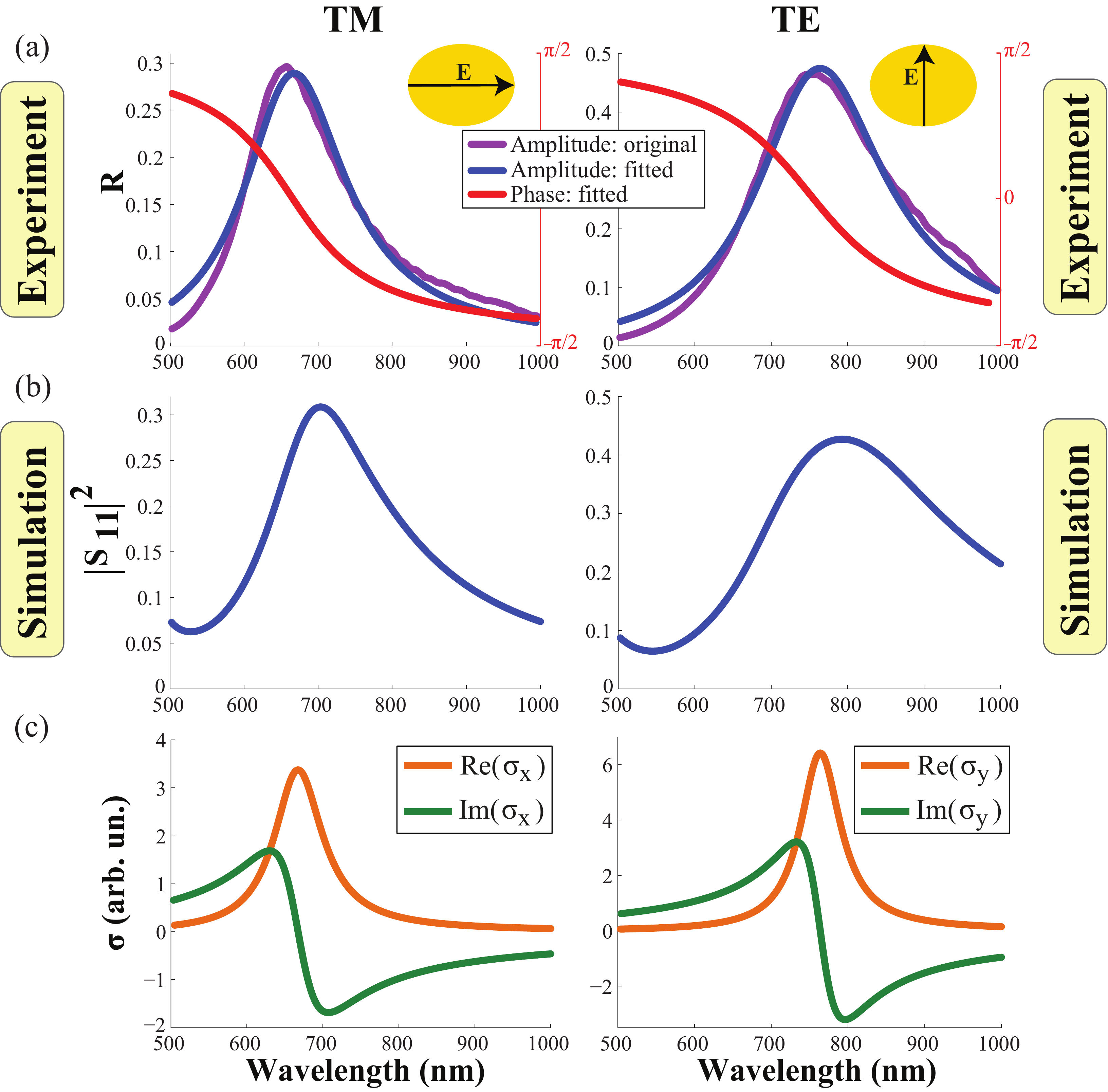}
  \caption{(a) Experimentally measured (purple) reflectance spectra of a metasurface for polarization along (left panel) and across (right panel) the long axis of the disk and its fitted amplitude (blue) and phase (red) by using Drude-Lorentz formula \eqref{sigma_drude}. (b) The reflectance of a metasurface calculated in CST Microwave Studio with six times increased losses of the gold. (c) Real (orange lines) and imaginary (green lines) parts of the TM- and TE-polarized components of the effective surface conductivity tensor extracted from the fitted reflectance by using Eq.~\eqref{fitting_formula}.}
 \label{fig:fitting}
\end{figure}

We consider a metasurface under consideration (Fig. 1) with the corresponding reflectance dispersion (See Fig. 2a). Then we perform the fitting of the Eq.~\eqref{fitting_formula} based on the least-squares method with Eq.~\eqref{sigma_drude} [See Fig.~8a]. To achieve the same order of the reflectance intensity in the simulation we increase the losses of the gold by six times (Fig.~8b). Finally, we obtain the dispersion of the surface conductivity for both polarizations from the fitting of the experimental data according to the Eq.~\eqref{sigma_drude} [See Fig.~8c]. By knowing the parameters of fitting formula Eq.~(s3) we can explicitly find the phase of the reflection coefficient from Eq.~(s4) [See Fig.~6a].

\section{Phase correction in frame of zero-thickness approximation}

Within zero-thickness approximation we substitute a plasmonic resonant metasurface of finite thickness $H$ by a two-dimensional layer with effective conductivity $\sigma$. The obvious question arises: at which distance $H_{\text{eff}}$ from the substrate we should dispose a two-dimensional layer? To define this distance we use a single criterion connected to the energy conservation law: 
\begin{equation}
\text{Re}(\sigma) > 0.
\label{criterion}
\tag{s7}
\end{equation}

The total phase of reflection coefficient (S-parameter) obtained by simulation in CST Microwave Studio is composed of two terms. The first one is an intrinsic phase
associated directly with the reflection from a metasurface $\varphi_0$, while the second term is an extrinsic part caused by the electromagnetic waves propagation from the excitation port to metasurface and back ($L_0$ is a distance between port and top of a metasurface). It is extremely important to define the extrinsic phase and make the appropriate phase correction in order to obtain the reflection coefficient related to the metasurface properties intrinsically. So, we can express the total phase $\varphi$ as 
\begin{equation}
\varphi = \varphi_0 + k_0 (2 L_0 + H_{\text{eff}}),
\tag{s8}
\end{equation}
 where $k_0 = n \omega/c$, $n$ is a refractive index of the super- or substrate.

Figure~9 shows the real parts of conductivity tensor components for the different locations of a two-dimensional layer (top, middle and bottom of a metasurface of finite thickness). One can see that this criterion is satisfied only, when a two-dimensional layer is disposed at the distance $H/2$ from the substrate.  

\begin{figure}[htbp]\centering
  \includegraphics[width=.88\linewidth]{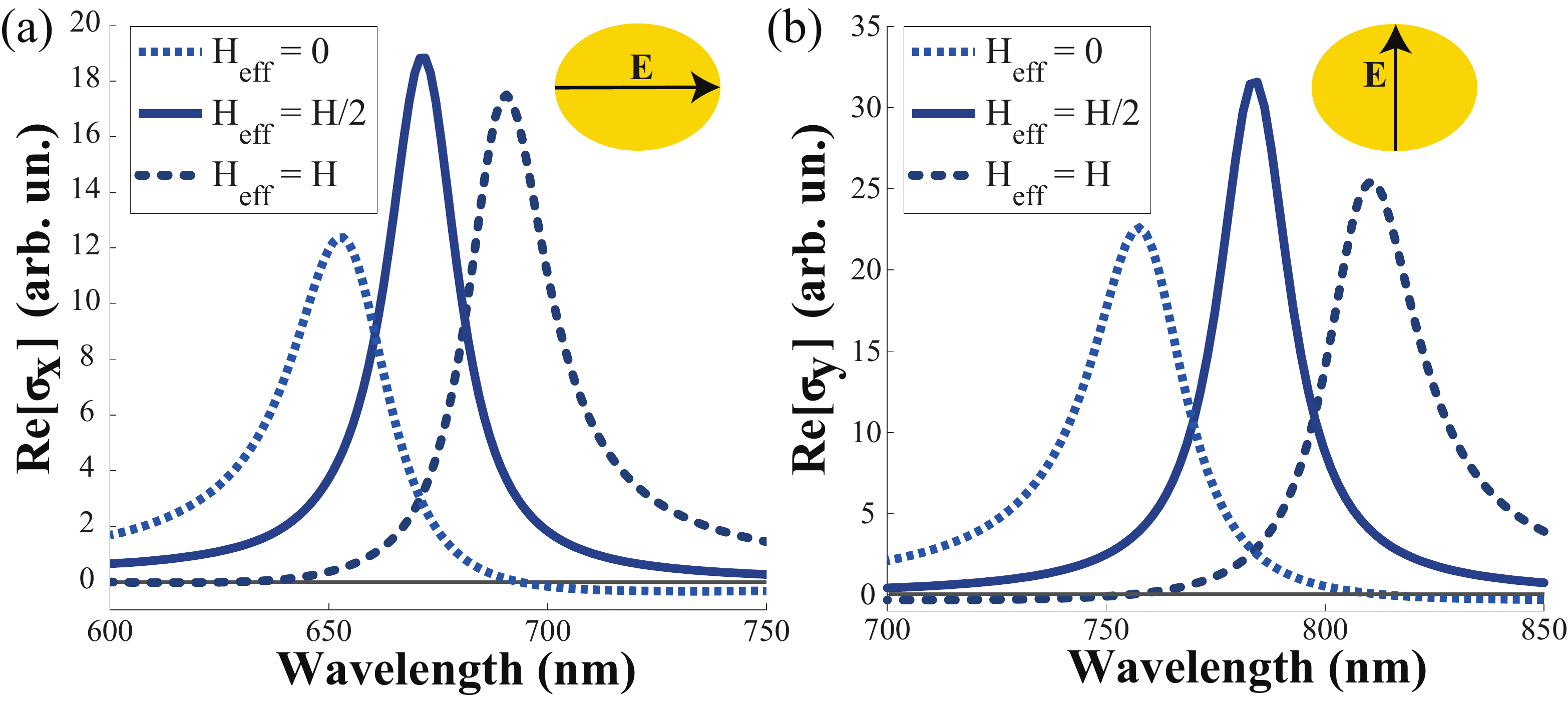}
  \caption{Real parts of conductivity for polarization along (a) and across (b) the long axis of the nanodisk at different distances between two-dimensional layer and substrate.}
 \label{fig:fitting}
\end{figure}

\section{Ewald summation for the 2D square lattice in 3D}

The lattice sums can be evaluated by using scalar Green's function in ${\bf r}$-space and ${\bf k}$-space: 
\begin{equation}
\begin{split}
& g(r) = \frac{e^{i k r}}{r} = - 4 \pi \int{\frac{d^3 k}{(2 \pi)^3}} \frac{e^{i {\bf k r}}}{k_0^2 - k^2}, \; r = |{\bf r - r_i}|, \\
& g_k({\bf r, r_i}) = \sum_i e^{i {\bf k r_i}} g({\bf r - r_i}) = - \frac{4 \pi}{V_0} \sum_{\bf b} \frac{e^{i {\bf (k+b) r}}}{k_0^2 - ({\bf k + b})^2}.
\end{split}
\tag{s9}
\end{equation}
Here $k_0$ is the wavevector of light in free space, ${\bf b}$ is the reciprocal wavevector of the structure, ${\bf r}$ is the excitation dipole or structural defect position and ${\bf r_i}$ is the position of $i$-th dipole of lattice structure. We consider the case ${\bf r} = (0,0,0)$.

Scalar Green's function (s9) can be divided into two parts by using Ewald summation with Ewald parameter $K$:
\begin{equation}
g_k({\bf r}) = \sum_{\bf b} e^{i {\bf (k + b) r}} f\left( {\bf k+b} \right) + \sum_i e^{i {\bf k r_i}} F\left( |{\bf r - r_i}|, K \right).
\label{g_k}
\tag{s10}
\end{equation}

For two-dimensional layer in the $xy$-plane with square lattice $a$ the first term is defined through the function
\begin{equation}
f\left( {\bf k+b} \right) = \frac{\pi}{a^2 \kappa_{{\bf k+b}}}  e^{-\kappa_{{\bf k+b}} |z|} \left[ 2 - \text{erf}\left( \frac{\kappa_{{\bf k+b}}}{2 K} - K z \right) - \text{erf}\left( \frac{\kappa_{{\bf k+b}}}{2 K} + K z \right) \right],
\tag{s11}
\end{equation}
where 
\begin{equation}
\kappa_{{\bf k+b}} = \sqrt{{\bf (k+b)^2} - k_0^2} \tag{s12},
\end{equation}
while the second term is expressed with the function
\begin{equation}
F\left( r, K \right) = \frac{\cos{\left( k_0 r \right)}}{r} - \frac{e^{i k_0 r}}{2 r} \text{erf} \left( K r + \frac{i k_0}{2 K} \right) - \frac{e^{-i k_0 r}}{2 r} \text{erf} \left( K r - \frac{i k_0}{2 K} \right).
\tag{s13}
\end{equation}

The overall sum should be not significantly dependent on the Ewald parameter $K$. This parameter is taken as $K \sim 1/a$, where $a$ is the lattice constant. So, we represent the Green's function as a sum of two contributions. The first term is calculated in real space, while the second is calculated in ${\bf k}$-space using Fourier transform: 

\begin{equation}
\hat{G}_k({\bf r}) = k_0^2 \left( 1 + \frac{1}{k_0^2} \text{grad div} \right) g_k({\bf r}) = \hat{G}_k^{(1)}({\bf r}) + \hat{G}^{(2)}({\bf r}).
\tag{s14}
\end{equation}

It significantly reduces calculation time, keeping accuracy to $10^{-4}$~\cite{capolino2007efficient}.

\end{suppinfo}

\bibliography{refer}

\end{document}